\documentclass[12pt]{article}
\usepackage{multirow}
\usepackage{graphicx}
\usepackage{cite}

\let\oldbibliography\thebibliography
\renewcommand{\thebibliography}[1]{%
  \oldbibliography{#1}%
  \setlength{\itemsep}{0pt}%
}
\newcommand\pubnumber{FERMILAB-CONF-14-437-E\\CDF note 11136}
\newcommand\pubdate{\today}

\def\Title#1{\begin{center} {\Large #1 } \end{center}}
\def\Author#1{\begin{center}{ \sc #1} \end{center}}
\def\Address#1{\begin{center}{ \it #1} \end{center}}

\newcommand\pubblock{\rightline{\begin{tabular}{l} \pubnumber\\
         \pubdate  \end{tabular}}}
\newenvironment{Abstract}{\begin{quotation}  }{\end{quotation}}
\newenvironment{Presented}{\begin{quotation} \begin{center} 
             PRESENTED AT\end{center}\bigskip 
      \begin{center}\begin{large}}{\end{large}\end{center} \end{quotation}}

\newcommand{\ttbar}{t \bar{t}}

\newcommand{\MET}{\mbox{$E\kern-0.50em\raise0.10ex\hbox{/}_{T}$}}
\newcommand{\met}{\mbox{$E\kern-0.50em\raise0.10ex\hbox{/}_{T}$}}

\newcommand{\afb}{A_{\mathrm{FB}}}

\newcommand{\afblep}{A_{\mathrm{FB}}^{\ell}}
\newcommand{\afbdeta}{A_{\mathrm{FB}}^{\Delta \eta}}
\newcommand{\afbtt}{A_{\mathrm{FB}}^{t\bar{t}}}

\newcommand{\Me}{\mbox{$E\kern-0.50em\raise0.10ex\hbox{/}$}}

\newcommand{\vtb}{\mathrm{V_{tb}}}

\begin{document}
\begin{titlepage}
\pubblock

\vfill
\Title{Top quark properties}
\vfill
\Author{Ziqing Hong\\(On behalf of the ATLAS, CDF, CMS and D0 collaborations)}
\Address{Mitchell Institute for Fundamental Physics and Astronomy, Texas A\&M University,
College Station, Texas 77843, USA\\
zqhong@fnal.gov}
\vfill
\begin{Abstract}
The top quark physics has entered the precision era. The CDF and D0 collaborations
are finalizing their legacy results of the properties of the top quark after the
shutdown of the Fermilab Tevatron three years ago. The ATLAS and CMS collaborations
have been publishing results from the LHC Run I with 7 TeV and 8 TeV proton-proton
collisions, with many more forthcoming. We present a selection of recent
results produced by the Tevatron and LHC experiments.
\end{Abstract}
\vfill
\begin{Presented}
XXXIV Physics in Collision Symposium \\
Bloomington, Indiana,  September 16--20, 2014
\end{Presented}
\vfill
\end{titlepage}
\def\thefootnote{\fnsymbol{footnote}}
\setcounter{footnote}{0}

\section{Introduction}

The top quark was first observed at the Fermilab Tevatron in 1995. Because the lifetime of the top quark is much shorter than
the hadronization time, it provides a unique opportunity to study a ``bare" quark. Due to
its large mass, the top quark may play a special role in understanding electroweak
supersymmetry breaking. 

The top quarks can be produced via both strong interactions and electroweak interactions
at the Tevatron and the LHC. The strong production of the top quarks is in the form
of $\ttbar$ pair production, with the quark anti-quark annihilation process dominating at the
Tevatron and the gluon-gluon fusion process dominating at the LHC. The electroweak production
of the top quarks is in the form of single top production. The $t$-channel diagram
dominates at both the Tevatron and the LHC, while the contributions from the $s$-channel
at the LHC and the $tW$-channel at the Tevatron are expected to be small.

This report summarizes a selection of recent results of the top quark properties reported by the CDF~\cite{CDFDetector} and D0~\cite{D0Detector} collaborations at the Tevatron experiment and the ATLAS~\cite{Aad:2008zzm} and CMS~\cite{Chatrchyan:2008aa} collaborations at the LHC experiment,
including measurements of the top quark mass, the mass differences between the top quark
and the top antiquark, the forward--backward asymmetry at the Tevatron and the charge
asymmetry at the LHC, the top spin correlations, the extraction of $|\vtb|$, and the top rare decays with flavor-changing neutral currents (FCNCs).

\section{Top quark mass}

The mass of the top quark is an important parameter of the standard model (SM). It is a
critical input for electroweak physics, and strongly related to vacuum stability.

\begin{figure}[hbtp]
\centerline{\includegraphics[width=0.9\columnwidth]{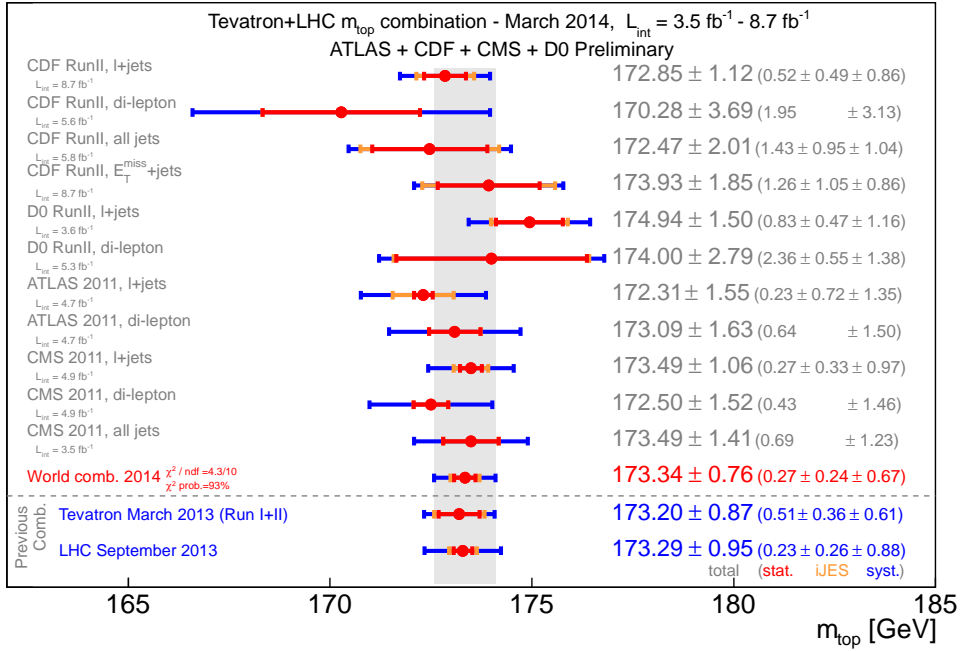}}
\caption{\label{topmasswc}First world combination of the top quark mass measurements.}
\end{figure}

The first world combination of the top quark mass was published in March 2014~\cite{ATLAS:2014wva}.
There are 17 measurements included in this combination, each of which are the most precise
measurements in each channel in each experiment. All the included measurements are based
on direct $\ttbar$ event reconstruction, and are all calibrated to the mass definition
used in Monte-Carlo (MC) generators. The uncertainty on the translation from this MC mass definition to a theoretically well defined mass is estimated to be on the order of 1~GeV~\cite{Moch:2014tta}. The combination yields a result of $m_{\mathrm{top}}=173.34\pm0.76~\mathrm{GeV}$, with a relative uncertainty of 0.44\%, as shown in Fig.~\ref{topmasswc}.

\begin{figure}[hbtp]
\centerline{\includegraphics[width=0.45\columnwidth]{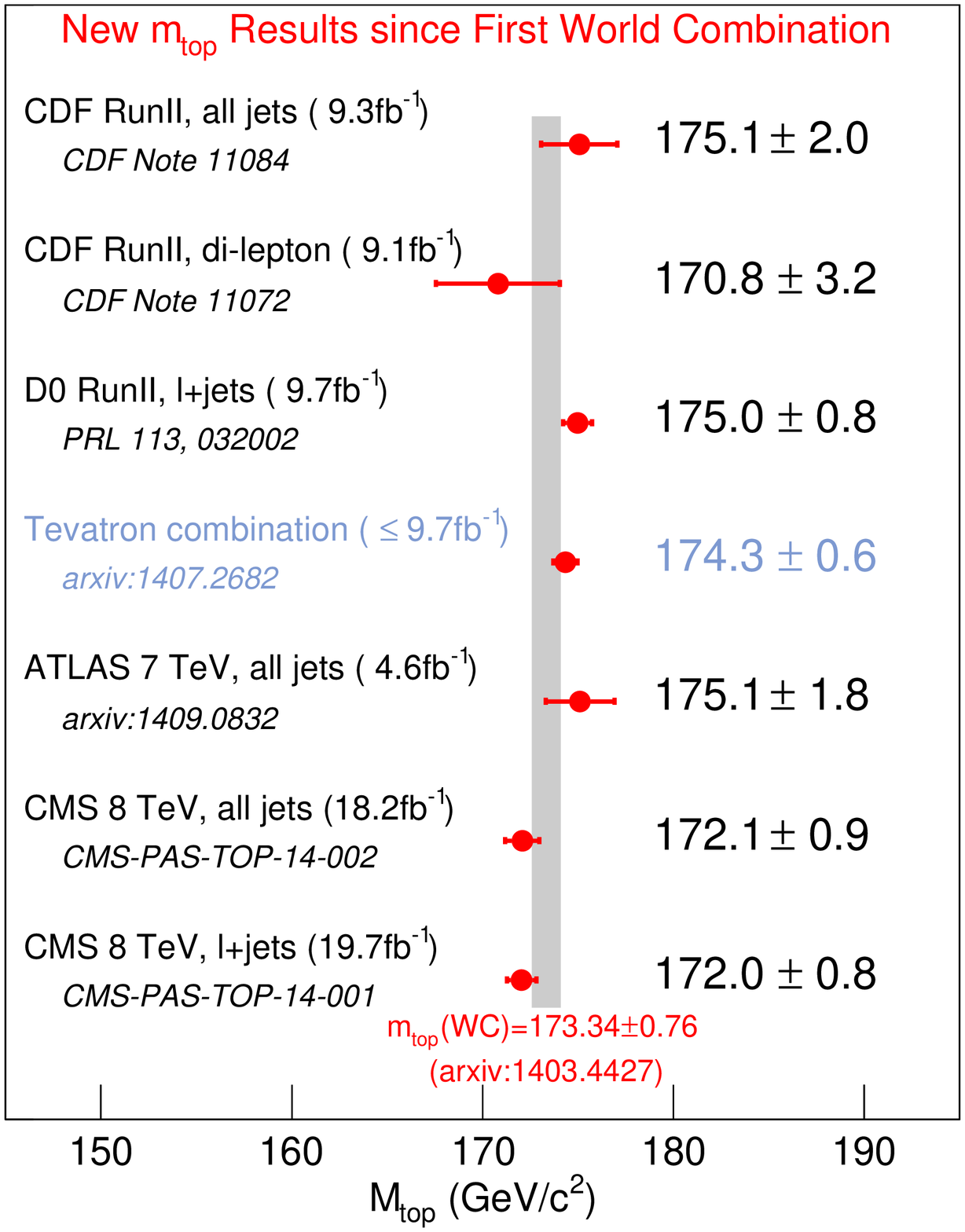}\includegraphics[width=0.45\columnwidth]{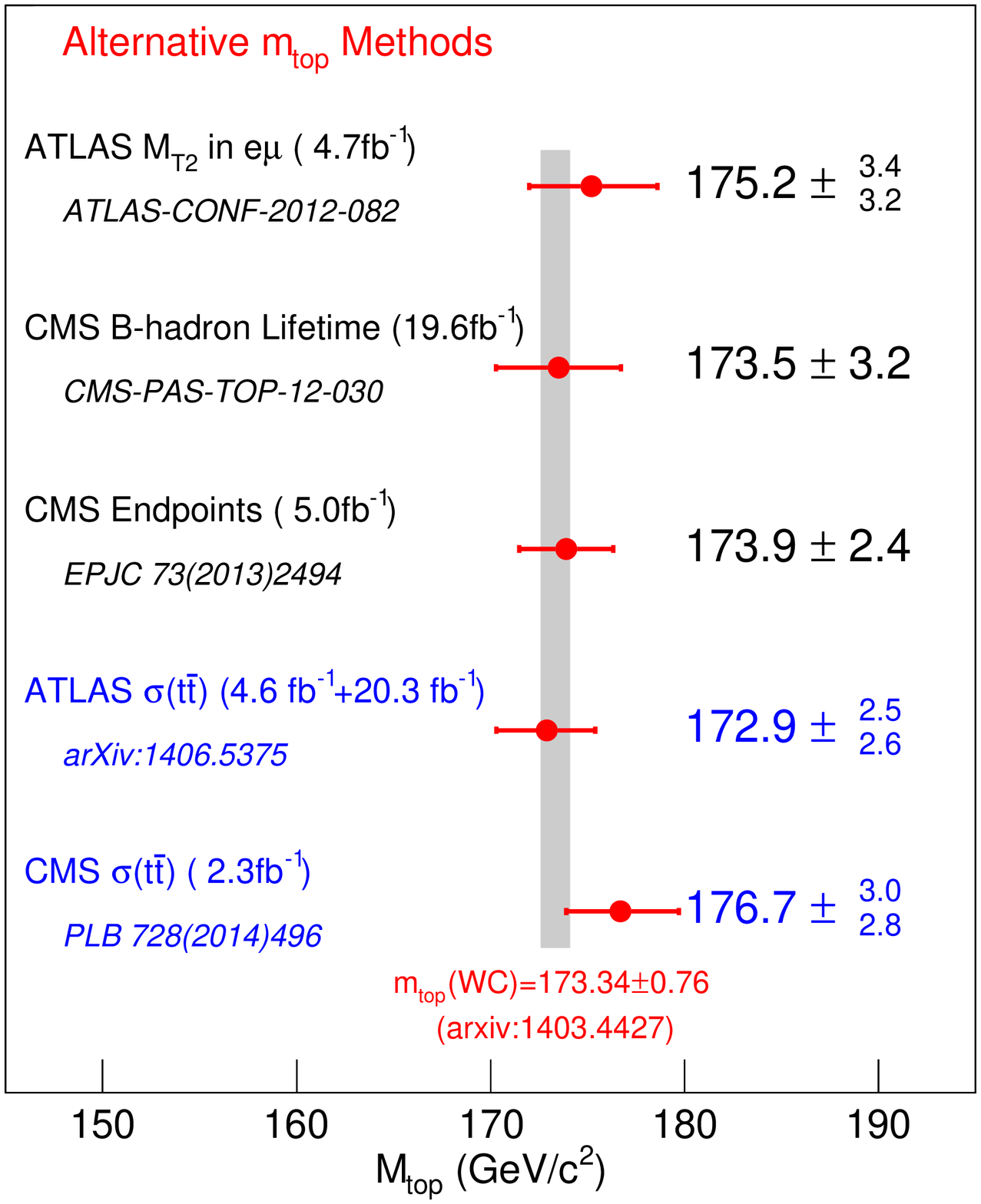}}
\caption{\label{topmassupdate}Updated top quark mass measurements after the first world combination (left) and top quark mass measurements with alternative methods from the LHC (right).}
\end{figure}

There are several updated measurements of the top quark mass after the first world combination, which are summarized in Fig.~\ref{topmassupdate}. The measurement in the lepton+jets channel from the D0 collaboration~\cite{Abazov:2014dpa} features the most precise single measurement to date. The Tevatron combination~\cite{Tevatron:2014cka} gives the smallest total uncertainty, with a result of $m_{\mathrm{top}}=174.34\pm0.64~\mathrm{GeV}$ and relative uncertainty of 0.37\%. The consistency among measurements is under study.

Alternative methods that do not involve direct $\ttbar$ event reconstruction have been developed at the Tevatron experiments and applied to measure the top quark mass by the LHC experiments with the results shown in Fig.~\ref{topmassupdate}. Since the lifetime of the B-hadrons is proportional to the top quark mass, the distance between the primary and the secondary vertices of a bottom jet has an approximately linear relation with the top quark mass~\cite{CMS:2013cea}, which allows for a measurement. Similarly, the $\mathrm{m_{T2}}$ variable, or the ``stransverse mass"~\cite{Lester:1999tx}, can be defined in events with two undetected particles, and as this variable has the property of establishing a lower bound on the parent particle's mass, which is the top quark mass in this case, a measurement can be made with it~\cite{ATLAS:2012poa}. The endpoint determinations in certain kinematic distributions are determined by the top quark mass, thus a measurement of the top quark can be based on these endpoints~\cite{Chatrchyan:2013boa}. Finally, the top quark mass can be extracted from the results of the top quark pair production cross section ($\sigma_{\ttbar}$), where the mass extracted from these measurements is the pole mass, rather than the MC-defined top quark mass. Note that while the alternative methods have the advantage that the uncertainties are less correlated with the measurements based on full $\ttbar$ reconstruction, none are as precise as the direct measurements.

\section{Mass difference between the top quark and the top antiquark}

The mass difference between the top quark and the top antiquark ($\Delta m = m_{t}-m_{\bar{t}}$) is predicted to be consistent with zero in the scenario of CPT conservation. 
Table~\ref{deltamtable} summaries the recent $\Delta m$ measurements from ATLAS, CDF, CMS and D0 collaborations. All results are consistent with zero.


\begin{table}[htbp]
\begin{center}
\begin{tabular}{l c l}
\hline
Experiment & $\Delta m$ (GeV) & Refs.\\\hline
CMS@ 8 TeV& $0.27\pm0.20(\mathrm{\normalsize stat})\pm0.12(\mathrm{\normalsize syst})$ &{\normalsize CMS-PAS-TOP-12-031}\\
CMS@ 7 TeV& $0.44\pm0.46(\mathrm{\normalsize stat})\pm0.27(\mathrm{\normalsize syst})$ &{\normalsize JHEP 06 (2012) 109}\\
ATLAS@ 7 TeV& $0.67\pm0.61(\mathrm{\normalsize stat})\pm0.41(\mathrm{\normalsize syst})$& {\normalsize PLB 728, 363 (2014) }\\
CDF RunII& $-1.95\pm1.11(\mathrm{\normalsize stat})\pm0.59(\mathrm{\normalsize syst})$& {\normalsize PRD 87 011101 (2013)}\\
D0 RunII& $0.8\pm1.8(\mathrm{\normalsize stat})\pm0.5(\mathrm{\normalsize syst})$& {\normalsize PRD 84, 052005 (2011)}\\
\hline
\end{tabular}
\caption{Summary of the recent $\Delta m$ measurements from ATLAS, CDF, CMS and D0. All results are consistent with zero.}
\label{deltamtable}
\end{center}
\end{table}

\section{Forward--backward asymmetry and charge asymmetry}

The forward--backward asymmetry ($\afb$) results at the Tevatron have shown potential anomalies for a number of years. Both the Tevatron and the LHC have been following up on them. The $\afb$ at the Tevatron and the charge asymmetry ($A_{C}$) at the LHC are both induced by the quark annihilation production mode. While there is no asymmetry arising from the leading order (LO) diagram, the next-to-leading order (NLO) QCD diagrams predicts that the top quark has a slightly larger probability of going in the direction of the quark in the initial state than the opposite direction.  The Tevatron has a proton-antiproton initial state, where quarks dominantly come from the protons and antiquarks dominantly come from the antiprotons. As a result, the top quark rapidity distribution is expected to have a few percent shift towards the proton direction (forward) and the top antiquark rapidity distribution in the opposite direction, thus resulting in the $\afb$. 


Two more variables that are sensitive to the top quark $\afb$ at the Tevatron are the leptonic $\afb$ defined with the pseudorapidity ($\eta$) of the charged leptons from the cascade decays of the top quarks in scenarios where at least one $W$ boson from the top quarks decays leptonically ($\afblep$), and the $\afb$ of the difference between the pseudorapidities of the two charged leptons from the cascade decays of the top quarks in scenarios where both $W$ bosons from the top quarks decay leptonically ($\afbdeta$).

\begin{figure}[hbtp]
\centerline{\includegraphics[width=0.6\columnwidth]{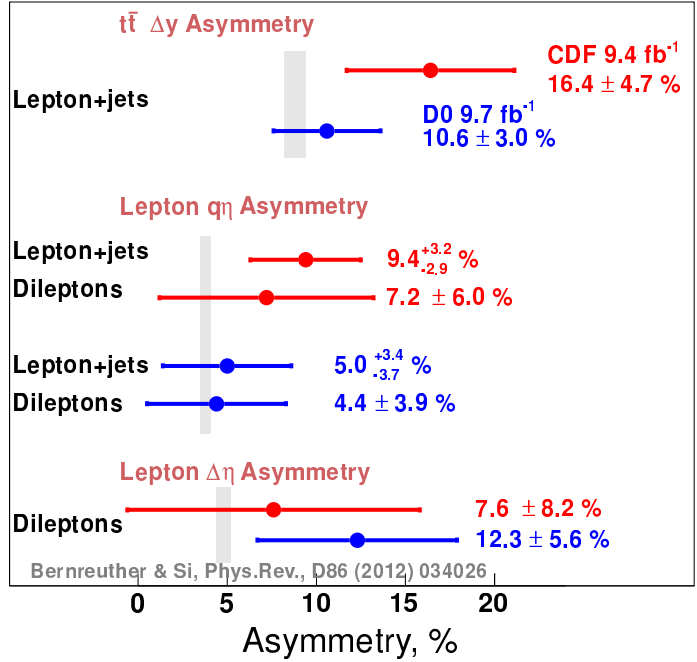}}
\caption{\label{fig:cdfd0comp}The results of $\afb$ from the CDF and D0 collaborations, comparing with the NLO SM calculations.}
\end{figure}

The results of $\afbtt$, $\afblep$ and $\afbdeta$ from CDF and D0 collaborations~\cite{Aaltonen:2012it,Abazov:2014cca,Aaltonen:2013vaf,
Aaltonen:2014eva,Abazov:2014oea,Abazov:2013wxa} are summarized in Fig.~\ref{fig:cdfd0comp}, comparing with the NLO SM calculation in Ref.~\cite{Bernreuther:2012sx}. The results from the CDF collaboration show tension with the NLO SM prediction, with the results from D0 collaboration consistent with both the results from the CDF collaboration and the NLO SM calculations. Note that all results are higher than the NLO SM calculations quoted.

A preliminary calculation of QCD at the next-to-next-to-leading order (NNLO) with LO electroweak corrections suggests $\afbtt \sim 0.10$~\cite{Mitov:CKM2014}. If this holds up, it means that the tension between the results from the CDF experiment and the predictions is no longer significant, and suggests that the QCD calculation at NNLO is needed for predictions for the kinematic distributions of the top quarks.

\begin{figure}[hbtp]
\centerline{\includegraphics[width=0.33\columnwidth]{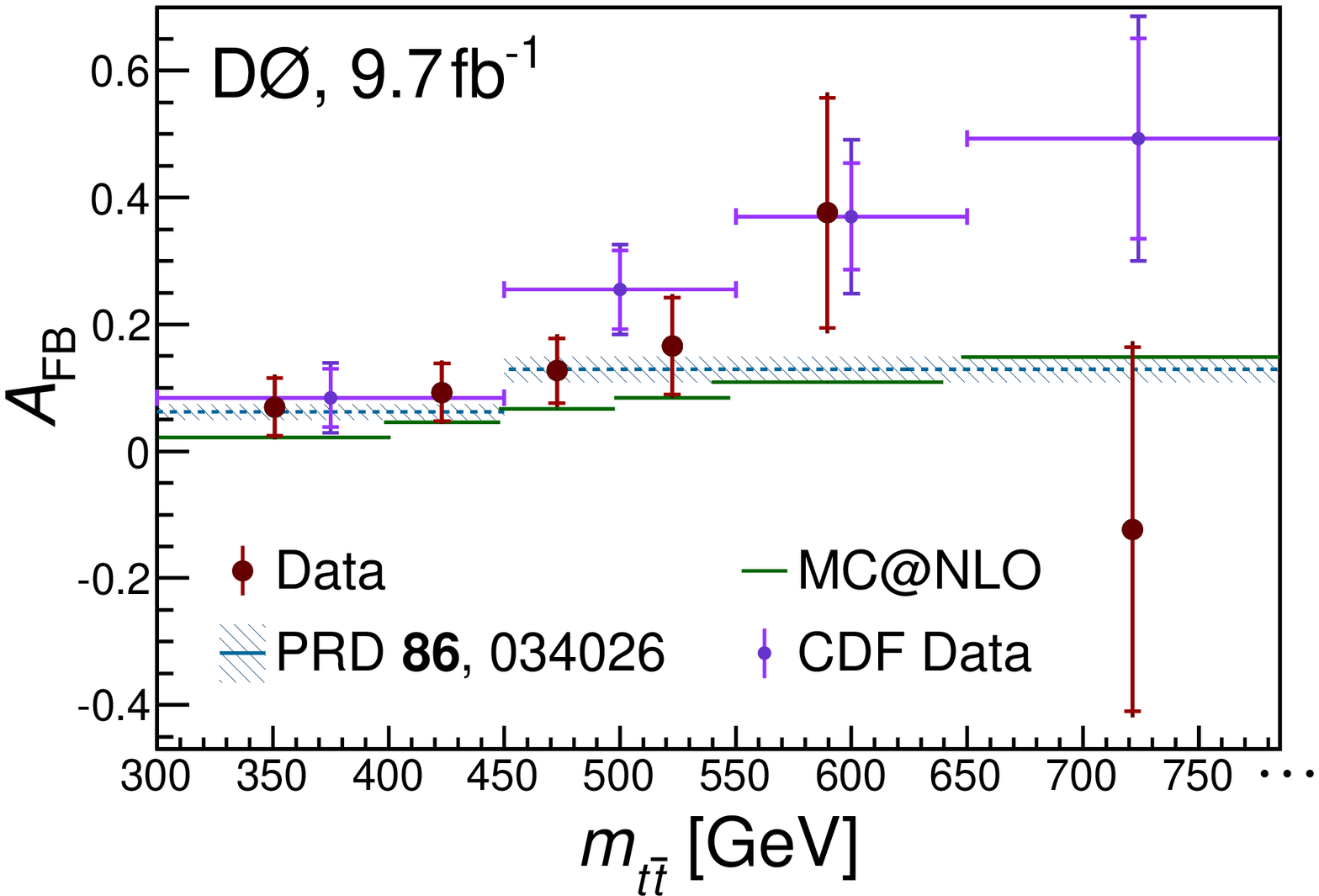}\includegraphics[width=0.33\columnwidth]{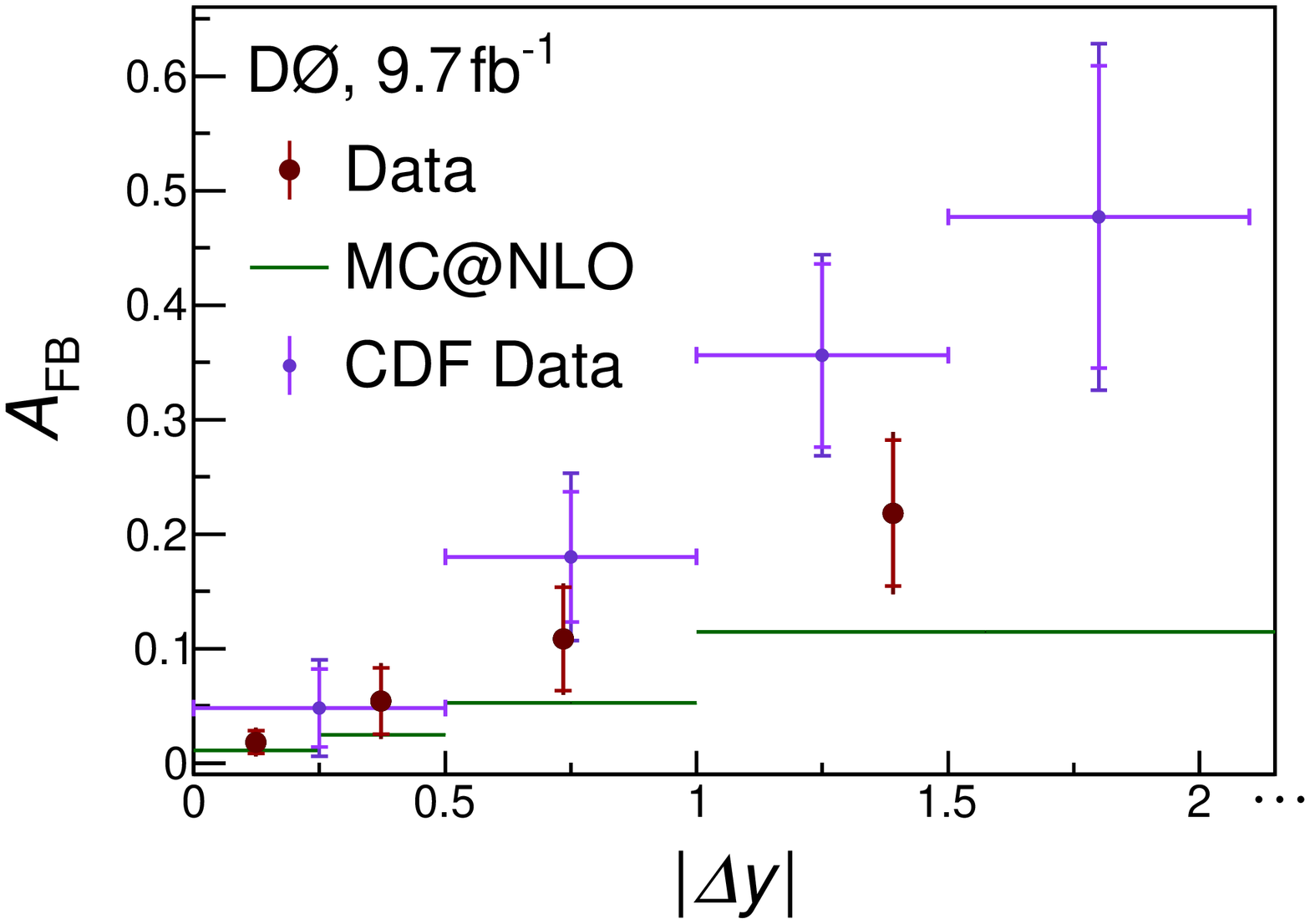}\includegraphics[width=0.33\columnwidth]{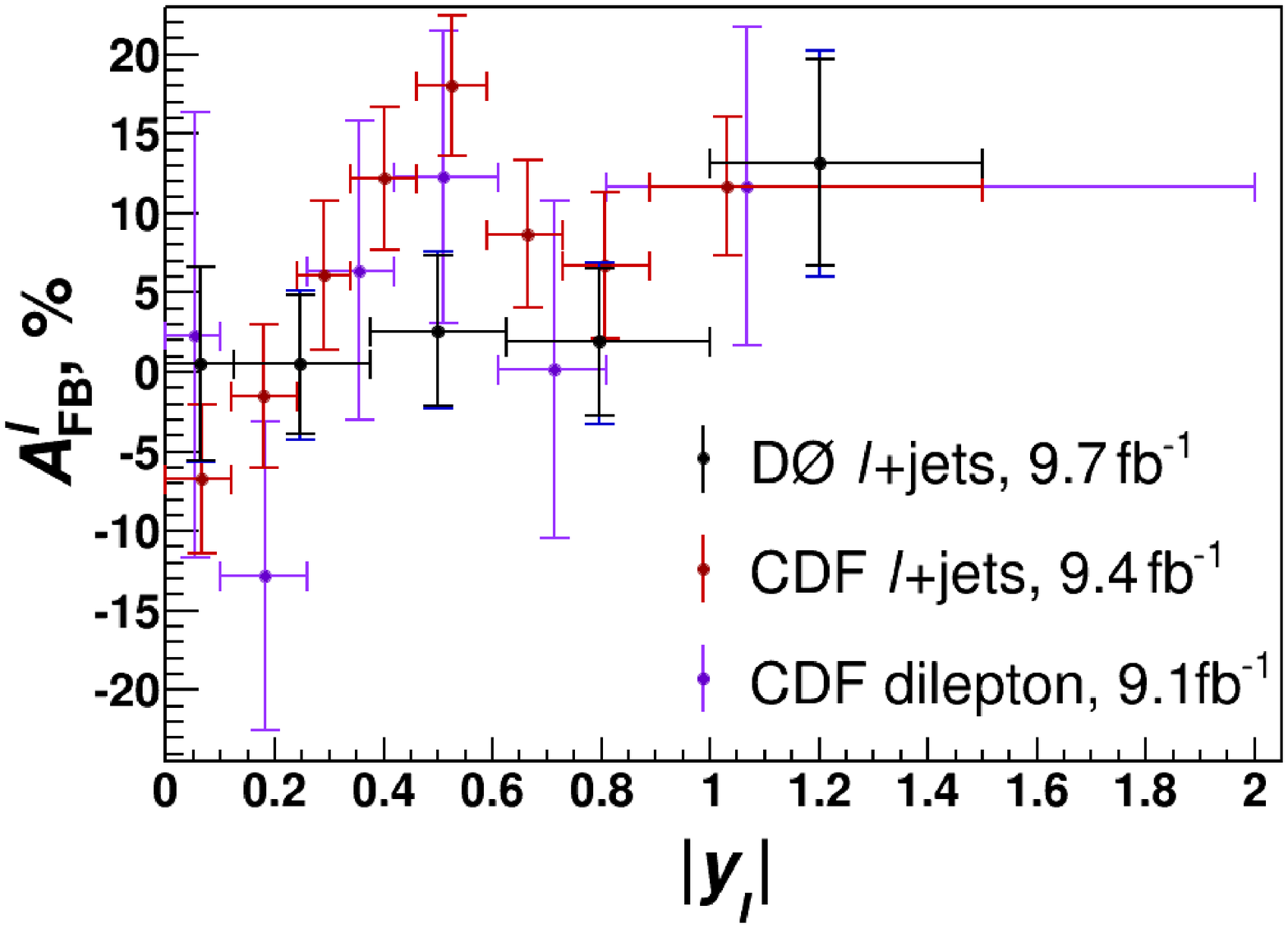}}
\caption{\label{fig:DIFAFB}The result of $\afbtt$ as a function of $m_{\ttbar}$ and $|\Delta y|$ and $\afblep$ as a function of $|y_{l}|$.}
\end{figure}

Both the CDF and D0 collaborations reported the differential $\afbtt$ as a function of $m_{\ttbar}$ and $|\Delta y|$ as well as the differential $\afblep$ as a function of $|y_{l}|$, shown in Fig.~\ref{fig:DIFAFB}. The D0 collaboration also reported the differential $\afblep$ as a function of $p_{T,\ell}$. The results from both collaborations show mostly good agreement with each other, with the regions with high $m_{\ttbar}$ and high $|\Delta y|$ under study.

The LHC experiment has followed up on this issue by measuring $A_{C}$. At the LHC, due to the proton-proton initial state, there is no predefined forward direction. Since there are both valance and sea quarks and only sea antiquarks in protons, the top quark rapidity distribution is expected to be broader than the top antiquark rapidity distribution, causing an effect characterized as $A_{C}$. Since the gluon fusion production mode dominates at the LHC, the $A_{C}$ is predicted to be even smaller. Measurements of both the inclusive~\cite{CMS:2014jua} (Table~\ref{table:INCAC}) and differential~\cite{Aad:2013cea,CMS:2013nfa,Chatrchyan:2014yta} asymmetries are made at the LHC with both 7~TeV and 8~TeV data. All results show agreement with the standard model predictions. We note that $\afb$ at the Tevatron and $A_{C}$ at the LHC can be related only in a model-dependent way.
\begin{table}
\begin{center}
\resizebox{\columnwidth}{!}{
\begin{tabular}{ccc}
\hline
Experiment&$A_{C}^{\ttbar}$&Refs.\\\hline
CMS&$0.004\pm0.010\mathrm{(stat)}\pm0.011\mathrm{(syst)}$&PLB 717, 129 (2012)\\
ATLAS&$0.006\pm0.010\mathrm{(stat)}\pm0.005\mathrm{(syst)}$&JHEP 1402 (2014) 107\\
ATLAS+CMS&$0.005\pm0.007\mathrm{(stat)}\pm0.006\mathrm{(syst)}$&ATLAS-CONF-2014-012/CMS-PAS-TOP-14-006\\\hline
Theory (NLO+EW)&0.0115$\pm$0.0006&JHEP 1201 (2012) 063\\\hline
\end{tabular}}
\caption{\label{table:INCAC}The results of the inclusive $A_{C}^{\ttbar}$ comparing with the SM prediction.}
\end{center}
\end{table}

\section{Spin correlation}

The spins of the top quark and the top antiquark in the top pair production are predicted to be correlated in the standard model, while it can be altered in new physics scenarios. Measurements have been done both at the Tevatron and at the LHC for the spin correlation of the top quark pairs. Three measurement methodologies can be applied in determining the spin correlation coefficients. One methodology is based on the azimuthal angles of the charged leptons from top quark decays~\cite{Aad:2014pwa}. 
The second methodology can be applied with the reconstruction of the top quark and antiquark momenta, with templates of the cosine of the lepton and bottom quark production angles in different spin correlation scenarios~\cite{CDF:10719}. The third methodology is based on matrix element technique~\cite{Abazov:2011gi}.

\begin{figure}[hbtp]
\begin{center}
\includegraphics[width=0.5\columnwidth]{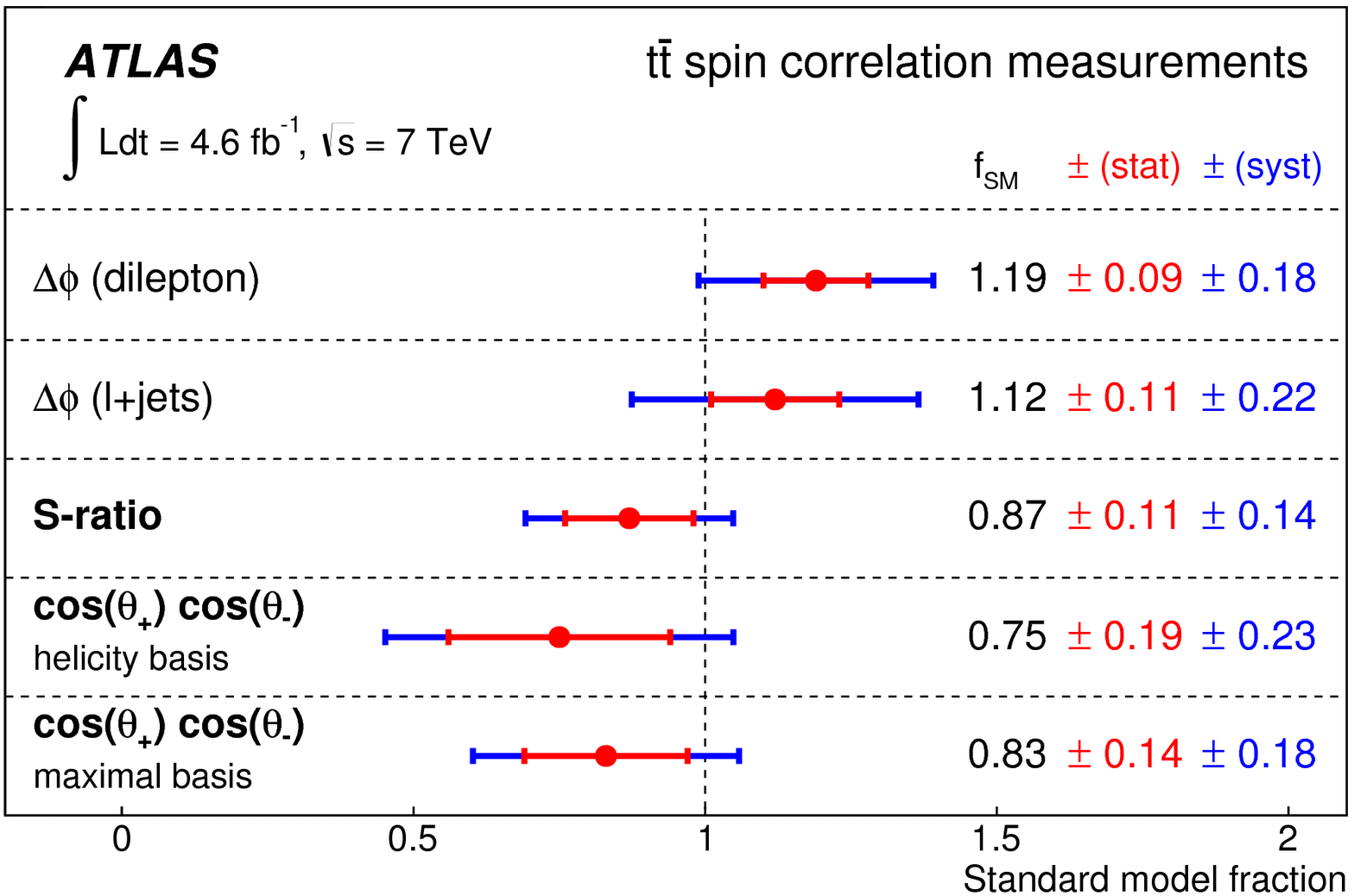}\\
\includegraphics[width=0.33\columnwidth, height=0.27\columnwidth]{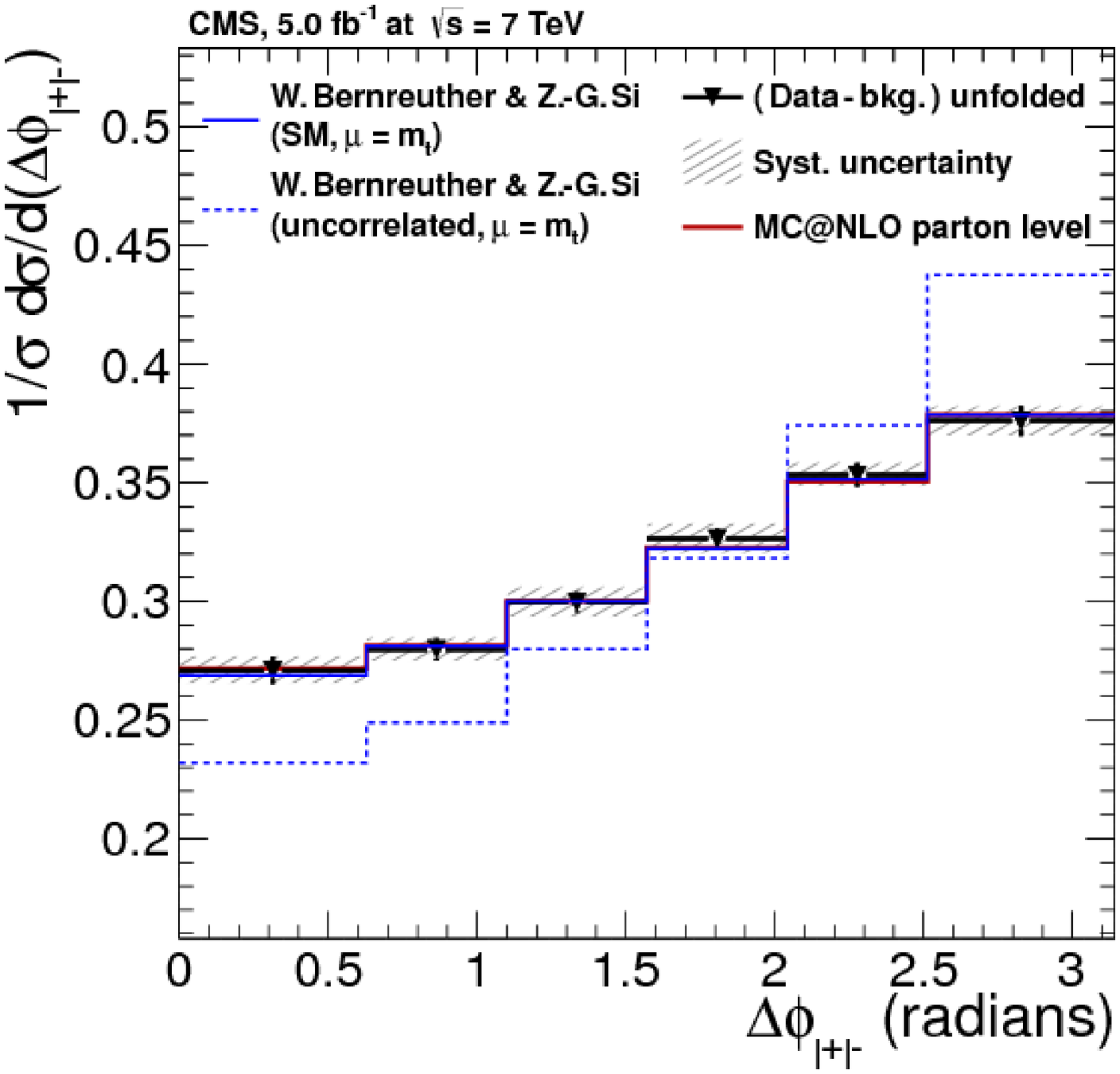}\hspace*{0.1cm}\includegraphics[ height=0.27\columnwidth,width=0.33\columnwidth]{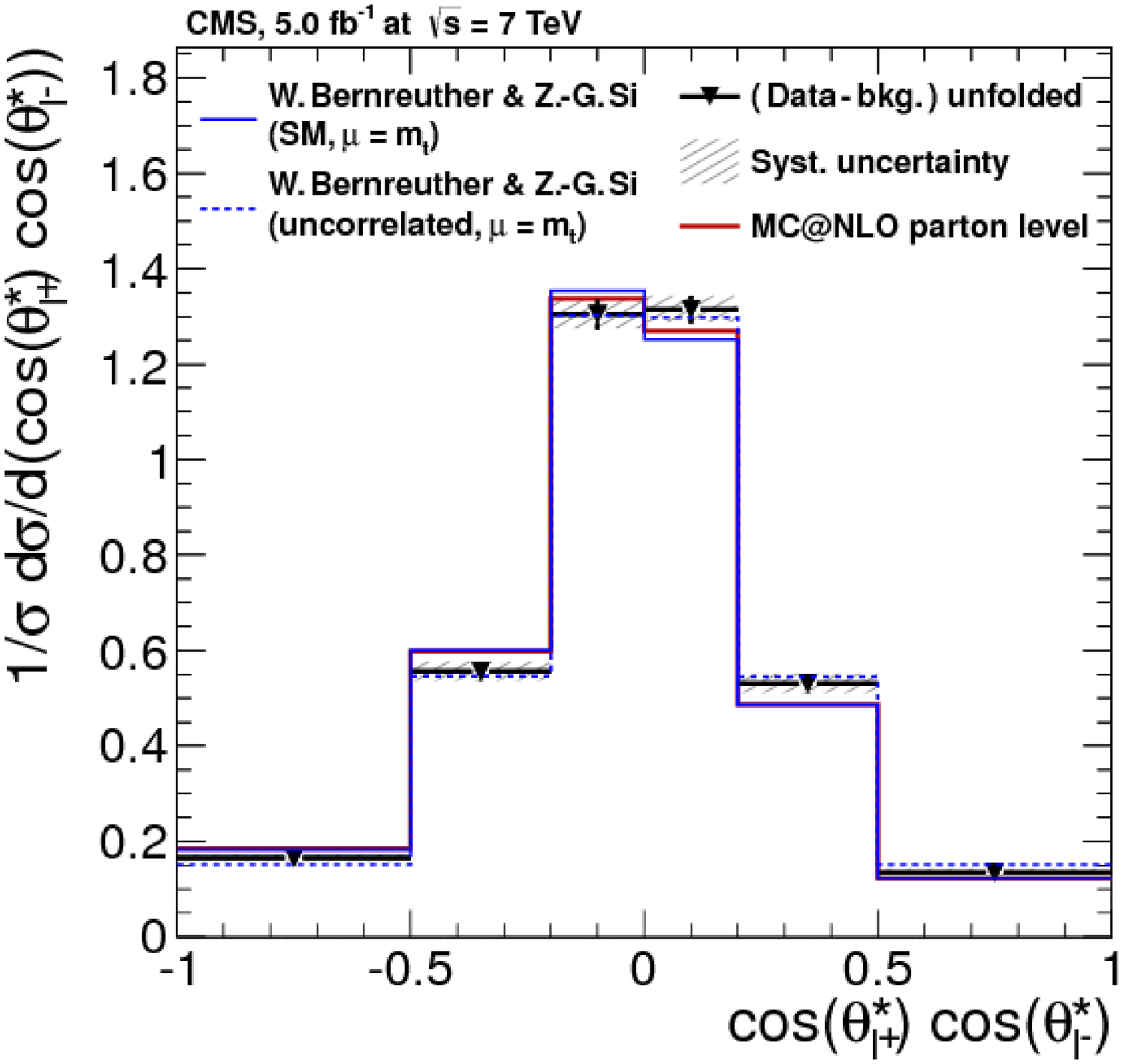}
\end{center}
\caption{\label{fig:LHCSpinCor}Results of the top spin correlation measurements at the LHC experiment. All results are consistent with the SM predictions.}
\end{figure}

The recent results of the top spin correlation measurements~\cite{CDF:10719,Abazov:2011gi} from the Tevatron experiment are summarized in Table~\ref{table:spincorTevatron}, together with the SM prediction~\cite{Bernreuther:2004jv}. Different methodologies were applied in measurements at the LHC~\cite{Aad:2014pwa,Chatrchyan:2013wua} and are summarized in Fig.~\ref{fig:LHCSpinCor}. All results are consistent with the SM predictions.
\vspace*{-\baselineskip}
\begin{table}[htbp]
\begin{center}
{\renewcommand{\arraystretch}{1.2}
\begin{tabular}{ccl}
 \hline
 Experiment&Spin correlation coefficient&Refs.\\\hline
 CDF&$0.042^{+0.563}_{-0.562}$&CDF Note 10719 (2011)\\
 D0 &$0.85\pm0.29$&PRL 108, 032004 (2012) \\\hline
 NLO SM&$0.78^{+0.03}_{-0.04}$&NPB 690, 81 (2004)\\
 \hline
\end{tabular}
}
\caption{Summary of the recent top quark spin correlation results from the Tevatron experiments. All results are consistent with the SM predictions.}
\label{table:spincorTevatron}
\end{center}
\end{table}

\section{$|\vtb|$ extractions}

The element $|\vtb|$ of the Cabibbo-Kobayashi-Maskawa (CKM) matrix is an important parameter of the SM. The value of $|\vtb|$ can be extracted both from measurements of the single top production cross section, assuming $|V_{\mathrm{tb}}|\gg |V_{\mathrm{td}}|~\mathrm{and}~|V_{\mathrm{ts}}|$, and from measurements of the branching ratio of $t\rightarrow Wb$, assuming in addition that the CKM matrix is a 3x3 unitary matrix. Table~\ref{table:TevatronVTB}~\cite{Aaltonen:2014ura,CDF:11033,Abazov:2013qka,CDF:2014uma,
Aaltonen:2013doa,Aaltonen:2014yua,Abazov:2011zk,Chatrchyan:2012zca,Chatrchyan:2014tua,Chatrchyan:2012ep,
Khachatryan:2014iya,Khachatryan:2014nda,Aad:2012ux,ATLAS:2014008,TheATLAScollaboration:2013fja} summarizes the results of $|\vtb|$ measurements from both the Tevatron and the LHC. All measurements lead to compatible results.

\begin{table}[htbp]
\begin{center}
\resizebox{\columnwidth}{!}{
\begin{tabular}{clccl}
\hline
Experiment & Channel & $|\vtb|$ & $|\vtb|$ lower limit @ 95\% C.L.&Refs.\\\hline
CDF&Single top (s+t) $\ell$+jets&$0.95\pm0.10$&0.78&arXiv:1407.4031\\
CDF&Single top $\met$+jets&-&0.63&CDF Note 11033 (2014)\\
CDF&Single top (s+t) combo&-&0.84&CDF Note 11033 (2014)\\
D0&Single top (s+t)&$1.12^{+0.09}_{-0.08}$&0.92&PLB 726, 656 (2013)\\
CDF&$\ttbar~\ell$+jets&$0.97\pm0.05$&0.89& PRD 87, 111101 (2013)\\
CDF&$\ttbar~\ell\ell$&$0.93\pm0.04$&0.85&PRL 112, 221801 (2014)\\
D0&$\ttbar~\ell$+jets/$\ell\ell$&$0.95\pm0.02$&0.96&PRL 107, 121802 (2011)\\
ATLAS&t-ch, 7~TeV&$1.13^{+0.14}_{-0.13}$&0.75&PLB 717, 330 (2012)\\
ATLAS&t-ch, 8~TeV&$0.97^{+0.09}_{-0.10}$&0.78&ATLAS-CONF-2014-007\\
ATLAS&tW-ch, 8~TeV&$1.10\pm0.12$&0.72&ATLAS-CONF-2013-100\\
CMS&tW-ch, 7~TeV&$1.010^{+0.163}_{-0.136}$&-&PRL 110, 02203 (2013)\\
CMS&tW-ch, 8~TeV&$1.030\pm0.126$&-&PRL 112, 231802 (2014)\\
CMS&t-ch, 7~TeV&$1.029\pm0.049$&-&JHEP 12(2013)035\\
CMS&t-ch, 8~TeV&$0.979\pm0.048$&-&JHEP 06(2014)090\\
CMS&t-ch, 7 \& 8~TeV comb.&$0.998\pm0.041$&-&JHEP 06(2014)090\\
CMS&$\ttbar~\mathrm{R_{b}}$, 8~TeV&1.007$\pm$0.016&-&PLB 736, 33 (2014)\\\hline
\end{tabular}}
\caption{Results of $|\vtb|$ measurements at the Tevatron and the LHC.}
\label{table:TevatronVTB}
\end{center}
\end{table}

\section{Search for flavor-changing neutral currents}

Processes involving flavor-changing neutral currents (FCNC) are heavily suppressed in the SM and result in the small branching ratios of the top rare decay modes. Potential enhancements of FCNC processes are possible in the presence of physics beyond the standard model~\cite{AguilarSaavedra:2004wm}. The limits on branching ratios of the top rare decay modes can be directly measured in $\ttbar$ channels as well as extracted from FCNC couplings at the production vertex involving single top production. Table~\ref{table:FCNC} summarizes the recent limits on the branching ratios of top rare decays~\cite{CMS:2014ffa,TheATLAScollaboration:2013vha,Chatrchyan:2013nwa,CMS:2014hwa,
Aad:2014dya,CMS:2014qxa}, comparing with the SM predictions and benchmark predictions from physics beyond the SM. The results are approaching the sensitivity to some plausible new physics models.

\begin{table}[htbp]
\begin{center}
\resizebox{\columnwidth}{!}{
 \begin{tabular}{c c c c c c c c c c}
 \hline
  & 7 TeV & 8 TeV & 7+8 TeV & 8 TeV & 7+8 TeV&8 TeV&7 TeV& \multirow{2}{*}{SM}& BSM\\
  & Single top&Single top&$\ttbar$ &Single top+$\gamma$&$\ttbar$&$\ttbar$&Single top + $Z$&&(MSSM)\\
  & CMS & ATLAS & CMS & CMS & ATLAS & CMS & CMS & &\\
  & CMS-PAS&ATLAS-CONF&PRL 112&CMS-PAS&JHEP 06&CMS-PAS&CMS-PAS&\multicolumn{2}{c}{hep-ph/0409342}\\
  &-TOP-14-007&-2013-063& 171802 (2014)&-TOP-14-003& (2014) 008&-HIG-13-034&TOP-12-021&\multicolumn{2}{c}{(2004)}\\
  \hline
  BR(t$\rightarrow$gu)& $<$ $3.6\cdot 10^{-4}$ & $<$ $3.1\cdot 10^{-5}$ & & & & &$<5.6\cdot10^{-3}$&$4\cdot 10^{-14}$&$2\cdot 10^{-6}$\\
  BR(t$\rightarrow$gc)& $<$ $3.4\cdot10^{-3}$ & $<$ $1.6\cdot 10^{-4}$ & & & & &$<7.1\cdot10^{-3}$&$5\cdot 10^{-12}$&$8\cdot 10^{-5}$\\
  BR(t$\rightarrow$Zq)& & &$<$ $5.0\cdot 10^{-4}$ & & & &$<5.1\cdot10^{-3}(Zu)/0.11(Zc)$&$1\cdot 10^{-14}$&$2\cdot 10^{-6}$\\
  BR(t$\rightarrow$u$\gamma$) & & & &$<$ $1.6\cdot 10^{-4}$& & &&$4\cdot 10^{-16}$&$2\cdot 10^{-6}$\\
  BR(t$\rightarrow$c$\gamma$) & & & &$<$ $1.8\cdot 10^{-3}$& & &&$5\cdot 10^{-14}$&$2\cdot 10^{-6}$\\
  BR(t$\rightarrow$ch) & & & & &$<$ $7.9\cdot 10^{-3}$ &$<$ $5.6\cdot 10^{-3}$ && $3\cdot 10^{-15}$&$1\cdot 10^{-5}$\\
  \hline 
 \end{tabular}}
\caption{Summary of recent limits on the branching ratios of top rare decays, compared with the SM predictions and benchmark predictions from physics beyond the SM.}
\label{table:FCNC}
\end{center}
\end{table}
\vspace*{-1.5cm}
\section{Conclusions}
The large mass and the short lifetime of the top quark makes it a fascinating particle with properties need to be further understood. A lot of measurements of the top quark production and properties have been done at both the Tevatron and the LHC. Top physics has entered the precision era, with some analyses already reaching the point where more precise QCD calculations are essential. Both CDF and D0 are finalizing their legacy results on top physics. More results are expected from LHC Run I data, with LHC Run II right around the corner.

\end{document}